\documentclass[aps,twocolumn,footinbib,showpacs]{revtex4}
\usepackage{epsf}
\let\iflong=\iffalse
\begin{document}
\title{Continuous Quantum Measurement and the Emergence of Classical Chaos%
       \vbox to 0pt{\vss
                    \hbox to 0pt{\hskip-50pt\rm LA-UR-99-3187\hss}%
                    \vskip 25pt}%
}
\author{Tanmoy Bhattacharya}
\email{tanmoy@lanl.gov}
\homepage{http://t8web.lanl.gov/t8/people/tanmoy/}
\author{Salman Habib}
\email{habib@lanl.gov}
\homepage{http://t8web.lanl.gov/t8/people/salman/}
\author{Kurt Jacobs}
\email{k.jacobs@lanl.gov}
\homepage{http://t8web.lanl.gov/t8/people/kaj/}
\affiliation{T-8, Theoretical Division, MS B285, Los Alamos
National Laboratory, Los Alamos, New Mexico 87545}

\begin{abstract}
We formulate the conditions under which the dynamics of a continuously
measured quantum system becomes indistinguishable from that of the
corresponding classical system.  In particular, we demonstrate that
even in a classically chaotic system the quantum state vector
conditioned by the measurement remains localized and, under these
conditions, follows a trajectory characterized by the classical
Lyapunov exponent.  
\end{abstract}
\maketitle


The emergence of classical chaos from quantum mechanics is probably
the most important theoretical problem in the study of the quantum to
classical transition.  Because of the absence of chaos in isolated
quantum systems~\cite{noqc} and the noncommutativity of the twin
limits $\hbar\rightarrow 0$ (the semiclassical limit) and
$t\rightarrow \infty$ (the late-time limit, necessary to describe
chaos), the fundamental mechanism of how classical chaos arises from
quantum mechanics
remains to be elucidated. While there has been much progress recently
in the development of sophisticated semiclassical methods for chaotic
dynamical systems \cite{semiclass}, attempts to unambiguously
characterize notions of chaos in the exact quantum dynamics
\cite{attempts} and to extract classical chaos as a formal
semiclassical limit have been less successful: a rigorous quantifier
of `quantum chaos' on par with the classical Lyapunov exponents has
yet to be found. And, since formal techniques have so far not succeeded
in extracting trajectories from isolated quantum systems, they have
not been able to explain the generation of chaotic time series in
actual experimental situations. The experimental state of the art has,
however, reached the stage where the quantum to classical transition
can now be probed directly~\cite{expts}.  So, it is crucial that one
understand the mechanism underlying this transition in order to
interpret existing results and design future experiments.

\iflong
Even though formal semiclassical limits of closed quantum systems with
few degrees of freedom are interesting from the point of view of
determining simple characteristics of these chaotic systems, real
experiments always deal with open quantum systems, {\it i.e.\hbox{}}
systems interacting with their environment.
\fi

As real experiments always deal with open systems, and the interaction
with the measuring apparatus necessary to deduce classical behavior
provides an irreducible disturbance on the free evolution of the
quantum system, the resulting decoherence and conditioned evolution
could play a crucial role in the emergence of the classical limit, and
of chaos, from the underlying quantum dynamics.  Indeed, some
qualitative results in this direction already exist \cite{hzs,qsd}.
In this Letter, we show that, even in the absence of any other
interaction with the environment, the theory of continuous quantum
measurements applied to the quantum dynamics of classically chaotic
systems provides a quantitatively satisfactory explanation of how
classical chaos, and Lyapunov exponents characterizing it, emerges
from quantum mechanics.

Open quantum systems are often studied by writing the evolution
equation for the reduced density matrix obtained by tracing over the
degrees of freedom in the environment.  
\iflong
Under this Master equation,
pure states evolve into mixed states and quantum interference
phenomena are suppressed by the averaging effect inherent in ignoring
the environment variables.
\fi
Although this process, called decoherence,
can be extremely effective in suppressing interference effects and
thereby making the quantum Wigner function approach the corresponding
classical phase space distribution function~\cite{hzs}, it does not
succeed in extracting localized `trajectories' from the quantum
dynamics.  Without the existence of such trajectories it is extremely
difficult, if not impossible, to rigorously quantify the existence of
chaos both mathematically and in actual experimental practice.
Since in order to extract classical trajectories systems must be
observed, 
\iflong
Furthermore, in real experimental settings, it is precisely some of
the environment variables, specifically, the recorded measurements,
which encode the information about the trajectories that the system
follows.  Hence, if quantum mechanics is to explain the emergence of
classical mechanics, 
\fi
one expects observed quantum systems to obey
classical dynamics in the macroscopic limit.

What is therefore desired is an unraveling of the Master equation
which provides a more detailed understanding of the trajectories
underlying the average system dynamics.  When these detailed
trajectories follow classical dynamics (albeit noisy), one can infer
that the average distributions generated by them also become
classical.  This then provides a `microscopic' understanding of the
quantum to classical transition demonstrated, {\em e.g.}, in
Ref.~\cite{hzs}. 

The first requirement in this program is to have a good model of
continuous quantum measurement.
Even though `continuous' measurement is always an idealization, real
experimental situations exist which approximate it extremely closely,
and simple models which correspond accurately to these processes have
now been developed \cite{cqm1,cqm1a,cqm2,cqm3}. 
These models show that as a necessary result of the information it
provides, continuous measurement produces and maintains localization
in phase space. 
On the other hand, the Ehrenfest theorem guarantees that
well-localized quantum systems effectively obey classical mechanics.
As the measurement process, in addition to localizing the state, also
introduces a noise in its evolution, to obtain classical mechanics one
must be in a regime in which the localization is sufficiently strong
and, yet, the resulting noise sufficiently weak.  We show that such a
regime exists, and is precisely the one which governs macroscopic
objects, {\it i.e.\hbox{}} $\hbar\ll S$, the action of the system.
In what follows, we refer to this regime as the classical regime.  Our
central result is that, once this regime is achieved, the localized
trajectories for the continuously observed quantum system obey the
classical dynamics (possibly chaotic) for that system driven by a weak
noise.  As a result, even at a finite but non-zero value of $\hbar$,
the quantum `trajectories' possess the same Lyapunov exponents as the
corresponding classical system.  As one goes deeper into the classical
regime with $\hbar\rightarrow 0$, one can make the noise progressively
smaller by optimizing the measurement, and, in the limit, the
intrinsic classical Lyapunov exponents are recovered.

In order for this mechanism to satisfactorily explain the
quantum-classical transition, the following conditions need to be
satisfied: (1) localization as discussed above, (2) suppression of
measurement noise, (3) the actual value of the measurement strength
should become irrelevant, and (4) the measurement record (i.e., the
actual results of the continuous measurement process), suitably
band-limited, should follow the classical trajectory. These conditions
are studied in more detail below.

We consider, for simplicity, a single quantum degree of freedom, with
position and momentum operators denoted by $X$ and $P$, evolving under
an unperturbed Hamiltonian $P^2/2m + V(X)$.  Quantum mechanics then
dictates the familiar Heisenberg equations of motion for these
operators: $\dot{X} = P/m$, and $\dot{P} = -\partial_X V(X) \equiv
F(X)$.  Except in the limit $\hbar \to 0$, a continuous observation
with finite measurement strength does not localize either the position
or the momentum completely.  Nevertheless, we can describe the state
of the particle in terms of the central moments of $X$ and $P$ and,
anticipating the limit, assign it to a point in phase space given by
the mean values $\langle X\rangle$ and $\langle P\rangle$.
\iflong
Provided the state {\em remains} localized, Ehrenfest's theorem
guarantees that these mean values obey classical equations of motion.
Conversely, for the closed nonlinear system the wavefunction spreads,
and as a consequence, the notion of an Ehrenfest trajectory rapidly
disappears even when the low order moments evolve similarly under
classical and quantum dynamics~\cite{hzs,corresp}.  For continuous
measurement to preserve classical dynamics, the localization it
produces must overcome the spreading caused by any nonlinearities.
\fi

The most natural measurement to use is a continuous measurement of
position, not only because this is often what is observed with
mechanical detectors, but also because {\em real} schemes for the
continuous measurement of position, considered in the field of quantum
optics, may be described very simply~\cite{measx}.  In addition, a
continuous measurement of position is an unraveling of the thermal
Master equation in the high temperature limit, so that results
demonstrated for this case also apply to decoherence due to a weakly
coupled, high temperature thermal bath.  We stress however, that we do
not expect the particular measurement model to effect the results
significantly; {\em any} measurement or interaction which produces a
localization in phase space should lead to classical behavior in
essentially the same manner.

Under continuous position measurement the evolution of the
wavefunction becomes stochastic. The stochastic master
equation for the density matrix $\rho (t)$, conditioned on the
measurement record $\langle X \rangle + \xi(t)$ with $\xi(t) \equiv
(8\eta k)^{-1/2}dW/dt$, is~\cite{measx}
\iflong
\begin{eqnarray}
|\tilde{\psi}(t + dt)\rangle &=& \left[ 1 - {1\over \hbar}\left(iH +
\hbar kX^2 \right) dt \right.\nonumber\\ 
&&\left. + \left(4k\langle X\rangle dt + \sqrt{2k} dW\right)X \right]
|\tilde{\psi}(t)\rangle ,
\label{nlsse}
\end{eqnarray}
\else
\begin{eqnarray}
\rho(t+dt) &=& \rho - (\frac{i}{\hbar} [H,\rho] - k [X,[X,\rho]]) 
                  \mathbin{} dt 
               \nonumber\\
           && {} + \sqrt{2\eta k} \mathbin{} ( [X,\rho]_+ - 
                2 \rho \mathop{\rm Tr} \rho X ) \mathbin{} dW \,,
\label{nlsse}
\end{eqnarray}
\fi
where $k$ is a constant specifying the strength of the measurement,
$\eta$ is the measurement efficiency and is a number between 0 and 1,
and $dW$ is a Weiner process, satisfying $(dW)^2=dt$.  When $\eta=1$,
the evolution preserves the purity of the state and can be rewritten
in a way which allows it to be understood as a series of diffuse
projection measurements~\cite{cqm2} on an unnormalized wavefunction
$\tilde\psi$:
\begin{eqnarray}
  |\tilde{\psi}(t + dt)\rangle = e^{-2kdt(X-(\langle X\rangle +
  \xi(t)))^2} e^{-iHdt/\hbar} |\tilde{\psi}(t)\rangle ,
\end{eqnarray}
where $\xi(t)$, the difference between $\langle X\rangle$ and the
measured value of the position, becomes a white noise in the limit
that $dt$ tends to zero.  
Under this continuous measurement process, the average values of
position and momentum evolve according to
\begin{eqnarray}
 d \langle X \rangle & = & (\langle P\rangle/m) dt + \sqrt{2k}V_x(t) dW \\
 d \langle P \rangle & = & \langle F(X)\rangle dt + \sqrt{2k} C_{xp}(t)
 dW ,
\label{seom}
\end{eqnarray}
where $V_x$ is the variance in position, and $C_{xp}$ is the
symmetrized covariance between $X$ and $P$~\cite{eomav}.  Thus the
effect of the measurement is to provide some zero-mean noise
proportional to the square root of the measurement strength $k$ and to
the width of the distribution.  It is important to note, however, that
this is just the first in a hierarchy of equations for the moments;
the equations governing the second moments contain terms depending on
higher moments, and so on up the hierarchy.  

Even though the structure of the hierarchy makes it almost impossible
to obtain analytic answers to questions regarding the behavior of the
variances, and resulting noise strength, which are the crucial
quantities determining the quantum to classical transition, we can,
nevertheless discuss the effect of varying $\hbar$ by truncating the
hierarchy at the second order, and looking at the steady state
solution for the variances.  These equations show that to maintain
enough localization to guarantee that, at a typical point on the
trajectory, $\langle F(x) \rangle \approx F(\langle x \rangle)$, as
required in the classical limit, the measurement strength, $k$, must
stop the spread of the wavefunction at the unstable
points\cite{footnote1}, $\partial_x F > 0$:
\begin{equation}
 8\eta k 
       \gg \frac{\partial_x^2 F}{F} 
             \sqrt{\frac{\partial_x F}{2 m}}\,.
\label{eq:klarge}
\end{equation}
Note that this condition is automatically satisfied for linear
systems, where quantum dynamics of the expectation values are
identical with classical evolution.

On the other hand, a large measurement strength introduces noise into
the trajectory.  If we demand that averaged over a characteristic time
period of the system, the change in position and momentum due to the
noise are small compared to those induced by the classical dynamics,
it is sufficient that, at a typical point on the trajectory, the
measurement satisfy 
\begin{equation}
\frac{2 \left\vert \partial_x F \right\vert}{\eta s} \ll \hbar k
   \ll \frac{\left\vert \partial_x F \right\vert s}{4}\,,
\label{eq:ksmall}
\end{equation}
where $s$ is the typical value of the action\cite{footnote2} of the
system in units of $\hbar$.  Obviously as $s$ becomes much larger than
$2\sqrt2\eta^{-1/2}$, this relationship is satisfied for an ever larger range
of $k$, and this defines the classical limit.

Finally, in experiments one usually considers the measurement record
itself rather than the estimated state of the system as we have done.
As measurement introduces a white noise, it is important to
investigate the condition under which the record tracks the estimate
faithfully.  If $\Delta t$ is the time over which the continuous
measurement is averaged to obtain the record, and we allow ourselves a
maximum of $\Delta x$ as the position noise, it is easy to see that
the measurement strength needs to satisfy
\begin{equation}
  8 \eta k > \frac{1}{\Delta t (\Delta x)^2}
\label{eq:kobs}
\end{equation}

\iflong
We notice that in the absence of any nonlinearity, the optimal
measurement induced localization is limited by the standard quantum
limit of $\sqrt{\hbar}$~\cite{Doherty}.  As the rate of spread of the
state due to the nonlinearity is itself proportional to the position
variance, any effect of nonlinearity decreases as $\hbar$ is made
smaller.  This suggests that a smaller measurement constant suffices
to maintain localization as $\hbar$ is made smaller, and the resulting
minimum width achievable also decreases.  As the noise depends on the
product of the measurement constant and the width, this in turn
implies that given a fixed noise threshold, a larger measurement
constant is possible as $\hbar$ decreases.  Thus, in the classical
regime one expects that the precise value of the measurement constant
becomes irrelevant to recovering the classical limit.
\fi

\begin{figure}
\leavevmode\epsfxsize=0.7\hsize\epsfbox{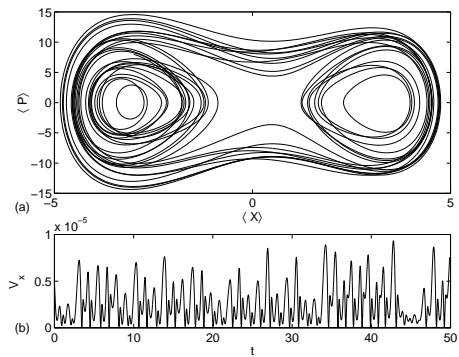}
\caption{
(a) The quantum trajectory in phase space, with
$\hbar=10^{-5}$ and $k=10^5$. (b) The position variance, $V_x$, as a
function of time.}
\label{fig1}
\end{figure} 

With this introduction, we consider, as an example, a bounded,
one-dimensional, driven system with the Hamiltonian,
\begin{equation}
H=P^2/2m + B X^4 - A X^2 + \Lambda X \cos(\omega t)~.
\label{lbham}
\end{equation}
with $m=1$, $B=0.5$, $A=10$, $\Lambda=10$, $\omega=6.07$.  This
Hamiltonian has been used before in studies of quantum chaos
\cite{linbal} and quantum decoherence \cite{hzs} and, in the parameter
regime used here, a substantial area of the accessible phase space is
stochastic.
\iflong
Using the calculational scheme described later, the 
finite-time classical Lyapunov exponent $\lambda \simeq 0.3 - 0.6$.
\fi
The numerical method used to solve Eq. (\ref{nlsse}) is a
split-operator, spectral algorithm implemented on a parallel
supercomputer.  

Simulations at various values of $\hbar$ confirm that as $\hbar$ is
reduced, both the steady-state variance, and the resulting noise (for
optimal measurement strengths) are reduced, as expected.  As the
dynamical time scale of this problem is $1 - 0.1$, we decide to
average the continuous observation record over a period of $0.01$.
Similarly, as the range of the motion covers distances of $O(10)$, we
demand that the position be tracked to an accuracy of $0.01$.  By
Eq.~\ref{eq:kobs}, this means we need $\eta k \sim O(10^{5})$ or
larger.  In our example, we choose the energy to be $O(10^2)$, and the
corresponding typical action turns out to be $O(10)$, and the typical
nonlinearity makes the rhs of Eq.~\ref{eq:klarge} $O(1)$. We see that
a choice of $\hbar = 10^{-5}$, $\eta = 1$ and $k = 10^5$, satisfies
all the constraints for a classical motion.  In
Fig.~\ref{fig1} we demonstrate that in this regime,
localization is maintained in spite of low noise.
Fig.~\ref{fig1}(a) shows a typical phase space trajectory, with the
position variance during the evolution, $V_x\equiv (\Delta X)^2$, plotted
in Fig.~\ref{fig1}(b). We find that the width $\Delta X$ is always
bounded by $3.4\times 10^{-3}$.  Furthermore, as is immediately
evident from the smoothness of the trajectory in Fig.~\ref{fig1}(a),
the noise is also negligible on these scales.

\begin{figure}[ht]
\begin{center}
\vbox{
\epsfclipon
\begin{center}
\leavevmode{\epsfxsize=0.65\hsize\epsfbox{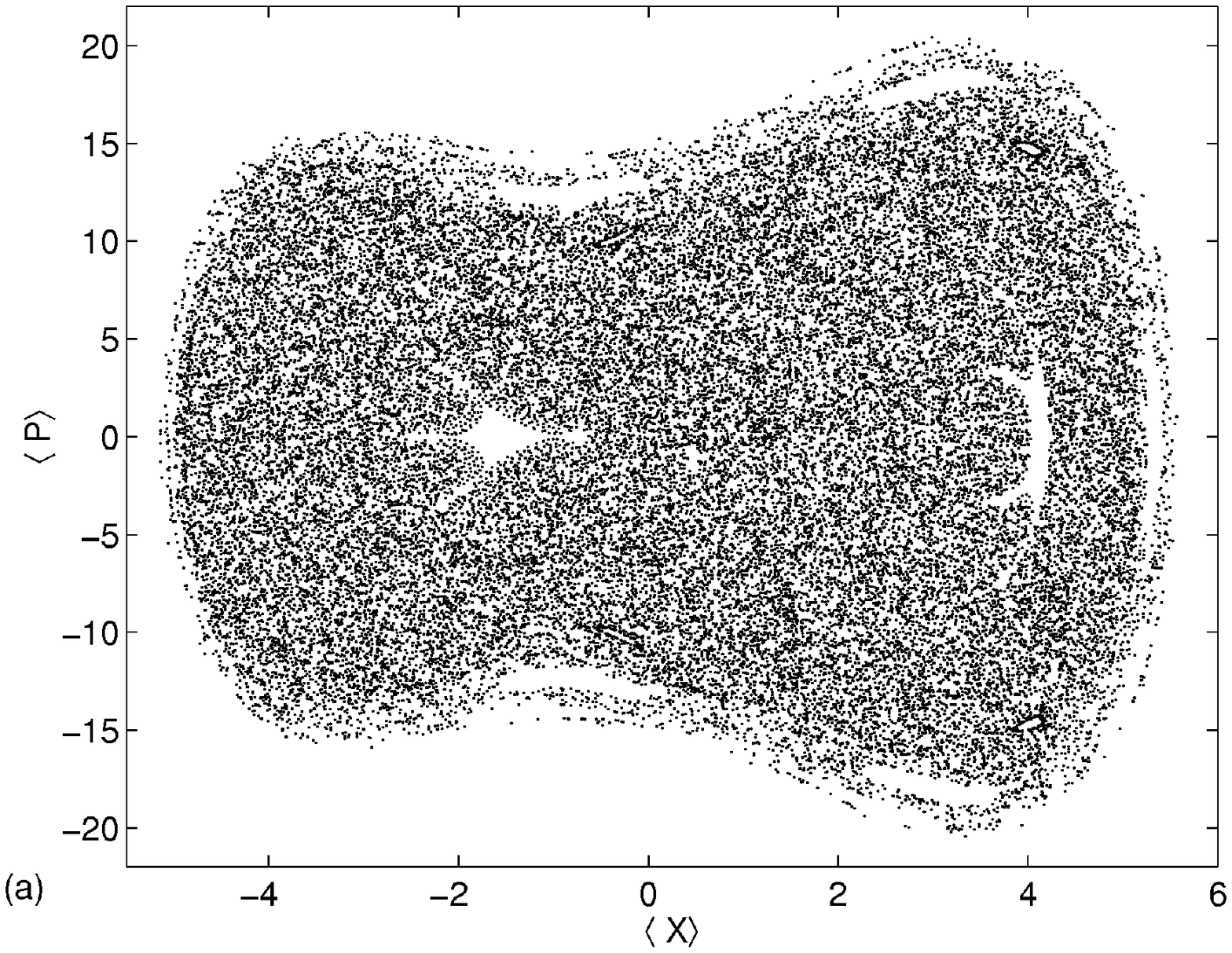}}
\relax\vskip 0.4cm\relax
\leavevmode{\epsfxsize=0.65\hsize\epsfbox{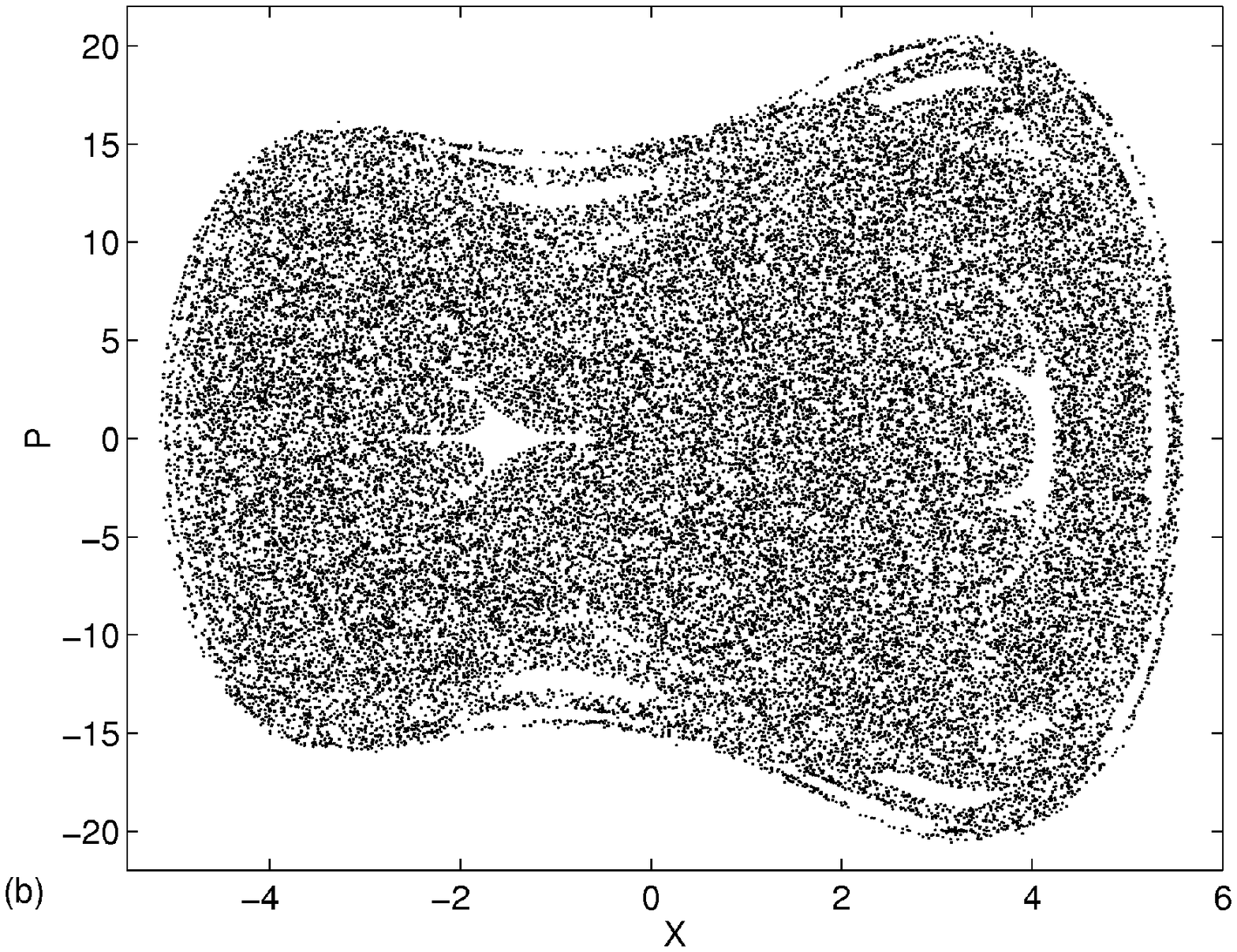}}
\end{center}
}
\end{center}
\caption{
(a) The quantum stroboscopic map with
$\hbar=10^{-5}$ and $k=10^5$. The figure is a pastiche from several
different runs with different initial conditions, for a total duration
of $39,000$ periods of the temporal drive. (b) The stroboscopic map
for the corresponding classical system, driven with a small amount of
noise.}
\label{fig2b}
\end{figure}

Thus, we find that continuous measurement can effectively obtain
classical mechanics from quantum mechanics.  
We substantiate this further by demonstrating that the trajectories we
obtain show the common signatures of classical chaos.  A direct way to
compare qualitatively the global nature of the quantum and classical
trajectories in phase space is to compare the stroboscopic maps (the
distribution of the locations of the system at a constant phase of the
driving term).  Fig.~\ref{fig2b} demonstrates the excellent
correspondence between the classical and quantum maps in this regard.

\begin{figure}
\leavevmode\epsfxsize=0.7\hsize\epsfbox{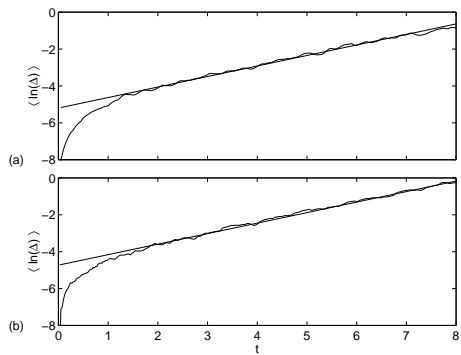}
\caption{
Lyapunov exponents, $\lambda$, calculated for (a)
the classical system driven with a small amount of noise, and (b) the
continuously observed quantum system, with $\hbar=10^{-5}$ and
$k=10^5$. The slope of the line drawn through the curves gives the
Lyapunov exponent, which in both cases is $\lambda=0.57(2)$.}
\label{fig3}
\end{figure}

On a more quantitative level, we now calculate the key characteristic
of chaos, the maximal Lyapunov exponent, $\lambda$, and compare it
against that of the classical system driven with a similar amount of
noise.  We start with the definition of $\lambda$: that for a chaotic
system the distance between two nearby trajectories, $\Delta(t)$,
evolves, on average, as $\ln \Delta(t) \sim \lambda t$, as long as
$\Delta(t)$ is small and $t$ is large.  To calculate this we take 10
fiducial trajectories (9 in the quantum case) starting at the point
(-3,8) and at 17 points along each trajectory, separated by time
intervals of 20 each, we obtain neighboring trajectories by varying
the noise realization.  The distance between the fiducial and these
neighboring trajectories is tracked for a time interval of 8.  The
values of $\ln \Delta(t)$ thus obtained are averaged over all the
instances and plotted versus $t$ in Fig.~\ref{fig3}, both for the
classical (with a small amount of noise) and the continuously measured
quantum systems.  For very small separations $\Delta$, the noise
dominates, which gives rise to an initial steep slope.  This is
followed by a linear region dominated by the Lyapunov exponent.
Eventually
$\Delta(t)$ becomes large and the curve flattens out.  The behavior of
the observed quantum and noisy classical systems are essentially
indistinguishable, and the Lyapunov exponent $0.57(2)$ is the
same for both. Performing the analysis with the classical system 
without noise, this time using 50 fiducial trajectories with initial 
points in a neighborhood of (-3,8), we obtain a Lyapunov exponent of 
$0.56(1)$, in agreement with the previous values.

After having demonstrated that in the classical regime, the
localization and low noise conditions are satisfied simultaneously, we
study the sensitivity to the measurement strength.  To this effect, we
vary $k$ between $2\times 10^{4}$ and $5\times 10^{5}$.  The Lyapunov
exponents remained unchanged within the quoted errors; only at
$k=5\times 10^{5}$ did the noise start to wash out the flat region of
the curve.


\begin{thebibliography}{99}
\bibitem{noqc} H.J. Korsch and M.V. Berry, Physica D {\bf 3}, 627
(1981); T. Hogg and B.A. Huberman, Phys. Rev. Lett. {\bf 48}, 711
(1982); R.L. Ingram, M.E. Goggin and P. W. Milonni, in {\em Coherence
and Quantum Optics VI}, edited by J.H. Eberly (Plenum, New York,
1990). 
\bibitem{semiclass} M.C. Gutzwiller, J. Math. Phys. {\bf 12}, 343
(1971); See also E.J. Heller and S. Tomsovic, Phys. Today {\bf 46}, 38
(1993).
\bibitem{attempts} See, {\it e.g.}, A. Peres, Phys. Rev. A {\bf 30},
1610 (1984); M. Toda and K. Ikeda, Phys. Lett. A {\bf 124}, 165
(1987); Y. Gu, {\em ibid} {\bf 149}, 95 (1990); R. Schack and
C.M. Caves, Phys Rev. E. {\bf 53}, 3257 (1996), Eprint: quant-ph/9506008.

\bibitem{expts} M. Brune, E. Hagley, J. Dreyer, X. Maitre, A. Maali,
C. Wunderlich, J. M. Raimond, and S. Haroche, Phys. Rev. Lett. {\bf
77}, 4887 (1996); C.J. Hood, M.S. Chapman, T.W. Lynn, and H.J. Kimble,
{\em ibid} {\bf 80}, 4157 (1998); H. Ammann, R. Gray, I. Shvarchuck,
and N. Christensen, {\em ibid} {\bf 80}, 4111 (1998); B.G. Klappauf,
W.H. Oskay, D.A. Steck, and M.G. Raizen, {\em ibid} {\bf 81}, 1203
(1998). 

\bibitem{hzs} S. Habib, K. Shizume, and W.H. Zurek,
Phys. Rev. Lett. {\bf 80}, 4361 (1998), Eprint: quant-ph/9803042.  
\bibitem{qsd} T.P. Spiller and J.F. Ralph, Phys. Lett. A {\bf 194},
235 (1994); T.A. Brun, I.C. Percival, and R. Schack, J. Phys. A {\bf
29}, 2077 (1996), Eprint: quant-ph/9509015.
\bibitem{cqm1} Early work includes A. Barchielli, L. Lanz and
G.M. Prosperi, Nuovo Cimento {\bf 72B}, 79 (1982); N. Gisin, Phys. Rev. Lett. {\bf
52}, 1657 (1984); L. Diosi, Phys. Lett. A {\bf 114}, 451 (1986);
for a review see, A. Barchielli, Int. J. Theor. Phys. {\bf 32}, 2221
(1993).
\bibitem{cqm1a} V.P. Belavkin and P. Staszewski, Phys. Lett. A {\bf 40}, 359 (1989).
\bibitem{cqm2} C.M. Caves and G.J. Milburn, Phys. Rev. A. {\bf 36},
5543 (1987).
\bibitem{cqm3} H. Carmichael, {\em An Open Systems Approach to Quantum
Optics} (Springer-Verlag, Berlin, 1993); H.M. Wiseman and
G.J. Milburn, Phys. Rev. A {\bf 47}, 642 (1993).
\iflong
\bibitem{corresp} L.E. Ballentine, Y. Yang, and J.P. Zibin,
Phys. Rev. A {\bf 50}, 2854 (1994); B.S. Helmkamp and D.A. Browne,
Phys Rev. E {\bf 49}, 1831 (1994); R.F. Fox and T.C. Elston, {\em
ibid} {\bf 49}, 3683 (1994); {\em ibid} {\bf 50}, 2553 (1994).
\fi
\bibitem{measx} A.C. Doherty and K. Jacobs, Phys. Rev. A {\bf 60},
2700 (1999), Eprint: quant-ph/9812004.
\bibitem{eomav} J.K. Breslin and G.J. Milburn, Phys. Rev. A {\bf
55}, 1430 (1997); J. Halliwell and A. Zoupas, Phys. Rev. D {\bf 52},
7294 (1995), Eprint: quant-ph/9503008.
\iflong
\bibitem{Doherty} A.C. Doherty, S.M. Tan, A.S. Parkins and D.F. Walls,
to appear in Phys. Rev. A, Eprint: quant-ph/9903030.
\fi
\bibitem{footnote1} Actually, if the nonlinearity is large on the quantum
scale, $\hbar \partial_x^2 F / F \gtrsim 4 \sqrt{\eta m \left\vert
\partial_x F\right\vert}$, $8\eta k$ needs to be much larger than
$(\partial_x^2 F(\langle x\rangle))^2 \hbar / 4 \sqrt\eta m F(\langle x
\rangle)^2$, irrespective of the sign of $\partial_x F$.  This does not
change the argument in the body of the paper.
\bibitem{footnote2} We are assuming that both $4 [m F^2 / (\partial_x
F)^2] \left\vert F/p \right\vert$ and $E \left\vert p/F \right\vert$
evaluated at a typical point of the trajectory are comparable to the
action of the system, and define this to be $\hbar s$.
\bibitem{linbal} W.A. Lin and L.E. Ballentine, Phys. Rev. Lett. {\bf
65}, 2927 (1990).

\end{thebibliography}
\end{document}